\documentclass[doublecol]{epl2}
\usepackage{amsbsy,amssymb,amsmath,bm,appendix}
\usepackage{graphicx,color}
\usepackage{amsfonts}
\usepackage{amssymb}
\usepackage{amsmath}
\usepackage{xspace,epsfig,float,multirow,subfigure,tabularx}

\pdfoutput=1

\title{Magnetic impurities make superconductivity in 3D Dirac semi-metal
triplet.}
\author{Baruch Rosenstein\inst{1,2} \and B.Ya. Shapiro \inst{3} \and %
Dingping Li \inst{4,5} \and I. Shapiro \inst{3} }

\institute{ \inst{1}Electrophysics Department, National Chiao Tung University, Hsinchu 30050, Taiwan, R. O. C\\
  \inst{2}Physics Department, Ariel University, Ariel 40700,Israel\\
 \inst{3}Physics Department, Bar-Ilan University, 52900 Ramat-Gan, Israel\\
 \inst{4}School of Physics, Peking University, Beijing 100871, China\\
\inst{5}Collaborative Innovation Center of Quantum Matter, Beijing, China}
\shortauthor{Baruch Rosenstein \etal}
\abstract{
 Conventional electron-phonon coupling induces either odd (triplet) or even
(singlet) pairing states in a time reversal and inversion invariant Dirac
semi - metal. In certain range of the chemical potential $\mu $ and
parameters characterizing the pairing attraction (effective
electron-electron coupling constant $\lambda $ and the Debye energy $T_{D}$)
the energy of the singlet although always lower, prevails by a very slim
margin over the triplet. This means that interactions that are small but
discriminate between the spin singlet and the spin triplet determine the
nature of the superconducting order there. It shown that in materials close
enough to the Dirac point ( $\mu \lesssim T_{D}$) magnetic impurities
stabilize the odd pairing superconducting state. }

\begin{document}

\maketitle

\textit{Introduction. }Recently solids with electronic states described by
the Bloch wave functions, obeying the "pseudo-relativistic" Dirac equation
(with Fermi velocity $v_{F}$ replacing the velocity of light) attracted
widespread attention. One outstanding example is graphene, a two-dimensional
(2D) hexagonal lattice made of carbon atoms. The 2D Dirac bi-spinor (spin in
this case is actually pseudospin/sublattice) incorporates excitations near
its $K$ and $K^{\prime }$ points in the Brillouin zone, so that the model is
in fact of the two band variety. Although a similar two band electronic
structure of bismuth was described by a nearly massless Dirac fermion in 3D,
this time caused by strong spin-orbit interactions, long ago \cite{Wolff}
(with spin replacing pseudospin), only recently several systems were
demonstrated to exhibit the 3D Dirac quasiparticles\cite{Liu,Potemski,Fang}.
Their discovery followed recent exploration of the topological band theory%
\cite{Zhang}.

A systematic proposal\cite{Kane} to make a 3D Dirac semi-metal is to close
the insulating gap by tuning a topological insulator towards the quantum
phase transition to trivial insulators led to their discovery. The time
reversal invariant 3D Dirac point in materials like $Na_{3}Bi$ was
theoretically investigated\cite{Wang13} and observed\cite{Liu}. A well known
compound $Cd_{3}As_{2}$ is a symmetry-protected 3D Dirac semi-metal with a
single pair of Dirac points in the bulk\cite{Fang}. Most recently
conductivity and magnetoabsorption of a zinc-blende crystal, $HgCdTe$ was
measured\cite{Potemski} and is in agreement with theoretical expectations in
Dirac semimetal \cite{Wan}. The discovery of the 3D Dirac materials makes it
possible to investigate their physics including remarkable electronic
properties. This is reach in new phenomena, not seen in 2D Dirac semi -
metals like graphene. Examples include the giant diamagnetism that diverges
logarithmically when chemical potential approaches the 3D Dirac point, slow
dynamics\cite{Wan}, linear in frequency AC conductivity that has an
imaginary part\cite{Wan}, quantum magnetoresistance showing linear field
dependence in the bulk\cite{Ogata}. Most of the properties of these new
materials were measured at relatively high temperatures. However some of
topological insulators and suspected 3D Dirac semi-metals exhibit
superconductivity at about the liquid $He$ temperature.

The well known topological insulator $Bi_{2}Se_{3}$ doped with $Cu$, becomes
superconducting at $T_{c}=3.8K$\cite{Ong}. When subjected to pressure\cite%
{pressureBiSe}, $T_{c}$ increases to $7K$ at $30GPa$. Quasilinear
temperature dependence of the upper critical field $H_{c2}$ that exceeds the
orbital and Pauli limits for the singlet pairing points to the triplet
superconductivity. The band structure of the superconducting compounds is
apparently not very different from its parent compound $Bi_{2}Se_{3}$.
Electronic-structure calculations of the compound under pressure\cite%
{pressureBiSe} reveal a single bulk three-dimensional Dirac cone like in $Bi$
with large spin-orbit coupling. Some experimental evidence point out to a
"conventional" phononic pairing mechanism. The reported values of
electron-electron due to phonons coupling constant $\lambda $ are probably
large with some reported values\cite{phononexp} well exceeding $\lambda =1$,
stronger than in good low $T_{c}$ superconducting metals. Theoretically the
spin independent part of the effective electron - electron interaction due
to phonons was studied\cite{DasSarma13}.\ In addition to $Bi_{2}Se_{3}$ and
similar compounds like $Bi_{2}Te_{3}$, the layered, noncentrosymmetric heavy
element $PbTaSe_{2}$ was found to be superconducting \cite{Cava}. Its
electronic properties like specific heat, electrical resistivity, and
magnetic-susceptibility indicate that $PbTaSe_{2}$ is a moderately coupled,
type-II BCS superconductor with large $\lambda=0.74$. It was shown
theoretically to possess a very asymmetric 3D Dirac point created by strong
spin-orbit coupling. If the 3D is confirmed, it might indicate that the
superconductivity is a conventional phonon mediated.

The case of Dirac semi-metal is very special due to strong spin dependence
of the itinerant electrons's effective Hamiltonian. It was pointed out\cite%
{FuBerg,Herbut} that in this case the triplet possibility can arise and
although the triplet gap is smaller than that of the singlet, the difference
sometimes is not large for spin independent electron - electron
interactions. Very recently the spin dependent part of the phonon induced
electron - electron interaction was considered\cite{DasSarma14} and it was
shown that the singlet gap is still larger than the triplet one. Another
essential spin dependent interaction is the exchange between itinerant
electrons and magnetic impurities\cite{expimp} like $Cr/Fe$ in $Bi_{2}Se_{3}$%
. Obviously it favors triplet, see Fig.1.
\begin{figure}[tbp]
\begin{center}
\includegraphics[width=8cm]{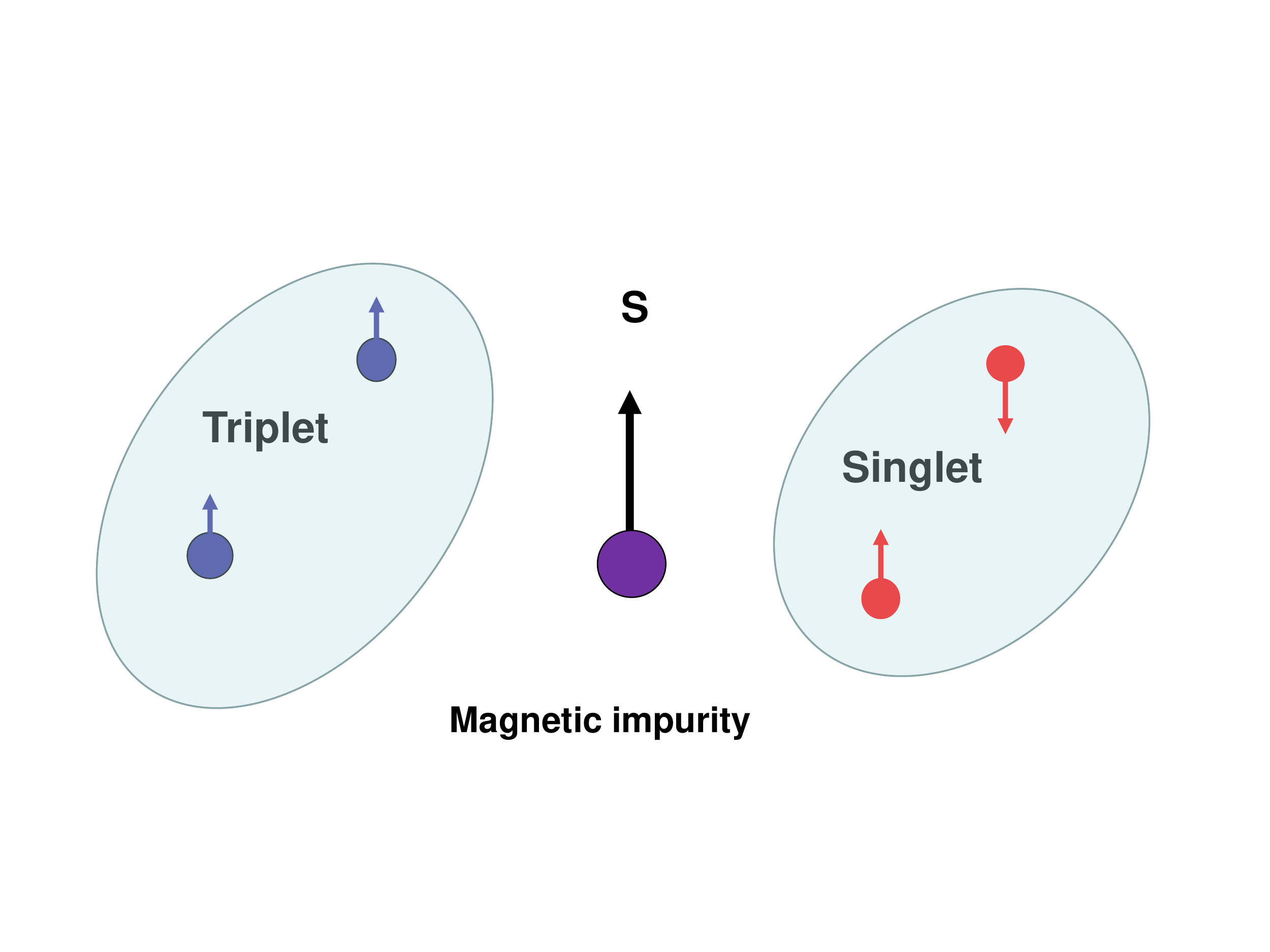}
\end{center}
\par
\vspace{-0.5cm}
\caption{Schematic picture of the impurity spin $\mathbf{s}$ interacting
with electrons composing a singlet (left) or triplet (right) Cooper pair. As
usual the magnetic impurities suppress the singlet, while is not pair
breaking for the triplet.}
\end{figure}
It therefore of importance to clarify theoretically two questions. (i) Does
a conventional phononic superconductivity exists in these materials with
just a minute density of states compared even with high $T_{c}$ cuprates
that apparently utilize very different pairing mechanism than phonons offer?
(ii) Is it possible that phonons in 3D Dirac materials lead to triplet
pairing that even becomes dominant under certain circumstances?

In the present letter we construct the theory of the superconducting
transition in 3D Dirac semi-metal at arbitrary chemical potential including
zero assuming the local (probably, but not necessarily, phonon mediated)
pairing. The possible pairing channels are classified in this rather unusual
situation using symmetries of the system. In contrast to the 2D case, the
odd parity (triplet) pairing is not only possible, but with small
concentration of magnetic impurities the odd parity is the preferred channel
taking over the more "conventional" even parity one.

\textit{Symmetry classification of pairing channels. }Electrons in 3D Dirac
semi-metal are described by field operators $\psi _{fs}\left( \mathbf{r}%
\right) $, where $f=L,R$ are the valley index for the left/right chirality
bands with spin projections taking the values $s=\uparrow ,\downarrow $.
These are combined into a four component bi-spinor creation operator, whose
index $\gamma =\left\{ f,s\right\} $ takes four values. The non-interacting
massless Hamiltonian with chemical potential $\mu $ reads\cite{Wang13},%
\begin{eqnarray}
K &=&\int_{\mathbf{r}}\psi _{\gamma }^{+}\left( -i\hbar v_{F}\nabla
^{i}\alpha _{\gamma \delta }^{i}-\mu \delta _{\gamma \delta }\right) \psi
_{\delta }\text{,}  \label{kinetic} \\
\text{ }\mathbf{\alpha } &=&\left(
\begin{array}{cc}
\mathbf{\sigma } & 0 \\
0 & -\mathbf{\sigma }%
\end{array}%
\right) =\left(
\begin{array}{cc}
0 & \mathbf{1} \\
\mathbf{1} & 0%
\end{array}%
\right) \mathbf{\gamma }\text{,}  \notag
\end{eqnarray}%
where $\sigma ^{i}$ and $\mathbf{1}$ are the Pauli and the unit matrices
respectively. We assume the time reversal, inversion and 3D rotational
symmetry that in particular requires an isotropic Fermi velocity. Electrons
interact electrostatically via the density - density potential. The
effective electron-electron interaction due to both electron - phonon
attraction and Coulomb repulsion (pseudopotential) can be taken local%
\begin{equation}
V_{eff}=\mathbf{-}\frac{g}{2}\int_{\mathbf{r}}\psi _{\gamma }^{+}\psi
_{\beta }^{+}\psi _{\beta }\psi _{\gamma }\text{.}  \label{phonons}
\end{equation}%
Unlike the free Hamiltonian $K$, Eq.(\ref{kinetic}), this interaction
Hamiltonian does not mix different spin components. Such a coupling
implicitly restricts the spin independent local interaction to be symmetric
under the band permutation and the additional term is not generated. A more
general case with additional independent term was considered in ref.\cite%
{FuBerg}. The strength of the phonon pairing depends on the cutoff: the
Debye temperature $T_{D}$.

Since we consider the local interactions as dominant, the superconducting
order parameter will be local $\widehat{M}=\int_{\mathbf{r}}\psi _{\alpha
}^{+}\left( \mathbf{r}\right) M_{\alpha \beta }\psi _{\beta }^{+}\left(
\mathbf{r}\right) ,$ where the constant matrix $M$ should be antisymmetric.
Due to the rotation symmetry they transform covariantly under infinitesimal
rotations generated by the spin rotation generators $S^{i}$, whose density
is
\begin{equation}
\mathbf{S}\left( \mathbf{r}\right) =\psi ^{+}\left( \mathbf{r}\right)
\mathbf{\Sigma }\psi \left( \mathbf{r}\right) ;\text{ \ \ \ \ \ \ }\mathbf{%
\Sigma =}\left(
\begin{array}{cc}
\mathbf{\sigma } & 0 \\
0 & \mathbf{\sigma }%
\end{array}%
\right) \text{.}  \label{spin}
\end{equation}%
The representations of the rotation group therefore characterize various
possible superconducting phases. Out of 16 possible matrices $M$ six are
antisymmetric. One finds one vector of the rotation group triplet $\mathbf{M}%
^{T}=\left\{ \gamma _{z},-\gamma _{x}\gamma _{y}\gamma _{z},\gamma
_{x}\right\} $ and three scalar multiplets: $M_{1}^{S}=i\alpha _{y};$ \ \ $%
M_{2}^{S}=i\Sigma _{y};$ \ \ $M_{3}^{S}=-i\gamma _{x}\gamma _{z}$ (see
Supplemental Materials (SM) for details\cite{SM}). In the odd parity
superconductivity state the rotational $O\left( 3\right) $ symmetry is
spontaneously broken, leading to weak ferromagnetism that has already been
considered (on level of the Ginzburg - Landau approach) for to heavy fermion
superconductor $UPt_{3}$\ \cite{Knigavko,Bel}. Which one of the condensates
is realized depends on energy determined by the interplay of the
interactions and disorder. Let us first consider clean homogeneous Dirac
semi-metal.

\textit{Singlet vs triplet.} The gap function for a channel $M$ can be
written\cite{AGD} as
\begin{equation}
\widehat{\Delta }_{\beta \gamma }=gF_{\beta \gamma }\left( r,\tau ;r,\tau
\right) =\Delta _{M}M_{\gamma \beta }\text{,}  \label{deltamatrix}
\end{equation}%
where $F\left( r,\tau ;r^{\prime },\tau ^{\prime }\right) \equiv
\left\langle T\psi _{\beta }\left( \mathbf{r,\tau }\right) \psi _{\gamma
}\left( \mathbf{r}^{\prime },\tau ^{\prime }\right) \right\rangle $ is the
anomalous Matsubara Green's function ($\Delta _{M}$ can be chosen real). The
Fourier transform of $F$ satisfies the matrix Gor'kov equation:%
\begin{equation}
F^{+}\left( \mathbf{p,}\omega \right) =-D^{t}\left( \mathbf{p,}\omega
\right) L^{+}\left( \mathbf{p,}\omega \right) G\left( \mathbf{p,}\omega
\right) \text{,}  \label{Gorkov}
\end{equation}%
where $D_{\gamma \beta }^{-1}=\left( i\omega -\mu \right) \delta _{\gamma
\beta }+v_{F}p^{j}\alpha _{\alpha \beta }^{j}$, while the Green's function $%
G $ obeys the Dyson equation,%
\begin{equation}
G^{-1}=D^{-1}+\widehat{\Delta }D^{t}\widehat{\Delta }^{\ast }\text{.}
\label{Dyson}
\end{equation}%
For the local phonon interaction, Eq.(\ref{phonons}), the operator
\begin{equation}
L_{ph}^{+}=-g\sum\nolimits_{\mathbf{q,\nu }}F^{+}\left( \mathbf{q,\nu }%
\right) ,  \label{Lph}
\end{equation}%
is independent of momenta and frequency and in view of Eq.(\ref{deltamatrix}%
) leads to an integrated form of the Gor'kov equation,
\begin{equation}
\widehat{\Delta }^{\ast }=-g\sum\nolimits_{\mathbf{p,}\omega }D^{t}\left(
\mathbf{p,}\omega \right) \widehat{\Delta }^{\ast }G\left( \mathbf{p,}\omega
\right) \text{.}  \label{gapeq}
\end{equation}%
The equation is solved with UV cutoff $T_{D}$ characterizing the
electron-phonon interaction along with the phonon-electron coupling $\lambda
=gD\left( \mu \right) =g\mu ^{2}/8\pi ^{2}v_{F}^{3}\hbar ^{3}$ for order
parameter $\Delta _{M}$ for all the channels. In Fig.2 the gap function for
singlet $S_{1}$, $\Delta _{S}$ in red, and $\Delta _{T}$ for triplet in blue
for chemical potentials $\mu =2T_{D}$ (left), $\mu =4T_{D}$(center) and $\mu
>>T_{D}$ (right, the BCS limit given in SM\cite{SM}). It turns out that $%
S_{3}$ is unstable, while $S_{1}$ and $S_{2}$ are degenerate.
\begin{figure}[tbp]
\begin{center}
\includegraphics[width=8cm]{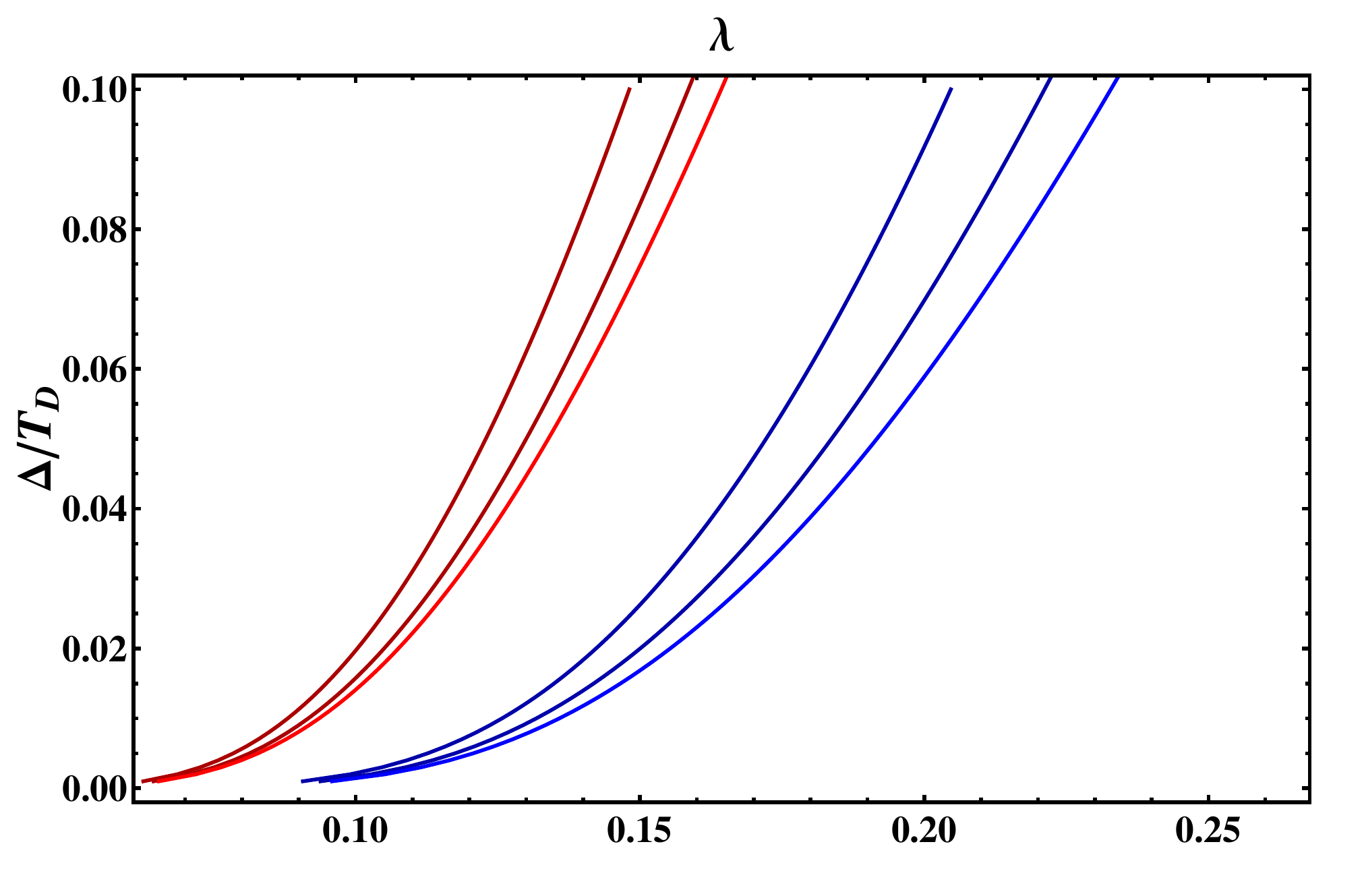}
\end{center}
\par
\vspace{-0.5cm}
\caption{Singlet (red) and triplet (blue) order parameters as function of
the phonon mediated effective electron-electron coupling $\protect\lambda $
at three values of chemical potential: $\protect\mu =2T_{D}$, $4T_{D}$ and
the BCS limit, $\protect\mu >>T_{D}$ (from left to right). }
\end{figure}
Examination of the Green's function reveals, see SM, that the spectrum of
excitations for \ singlet is rotational invariant, while the one of triplet
has two nods. One finds that the singlet has larger gap function for $\mu >1$%
, although at small chemical potential and large coupling the triplet order
parameter actually is a bit higher than that of the singlet. The value of
the gap function itself does not define which channel is stable, so we have
calculated energy densities via momentum space Green's function for all the
channels,%
\begin{eqnarray}
E_{S,T} &=&\int_{\Delta =0}^{\Delta _{S,T}}\frac{d\left( 1/g\left( \Delta
\right) \right) }{d\Delta };  \label{energies} \\
\frac{1}{g\left( \Delta \right) } &=&-\frac{1}{2}\sum\nolimits_{\mathbf{%
p,\omega }}\text{Tr}\left( M^{S,T}D^{t}\left( \mathbf{p},\omega \right)
M^{S,T}G\left( \mathbf{p},\omega \right) \right) \text{,}  \notag
\end{eqnarray}%
see SM for details\cite{SM}. Limiting cases of BCS when $\mu >>T_{D}$ can be
done analytically, see SM, while experimentally relevant (see below)
chemical potentials $\mu =8T_{D}$ and $12T_{D}$ are given for wide range of
couplings in Fig.3. Triplet (blue line) has always higher energy than
singlet although at $\mu <T_{D}$ energies of triplet and singlet are close
despite the fact that $\Delta _{T}>\Delta _{S}$, see SM\cite{SM}.
\begin{figure}[tbp]
\begin{center}
\includegraphics[width=8cm]{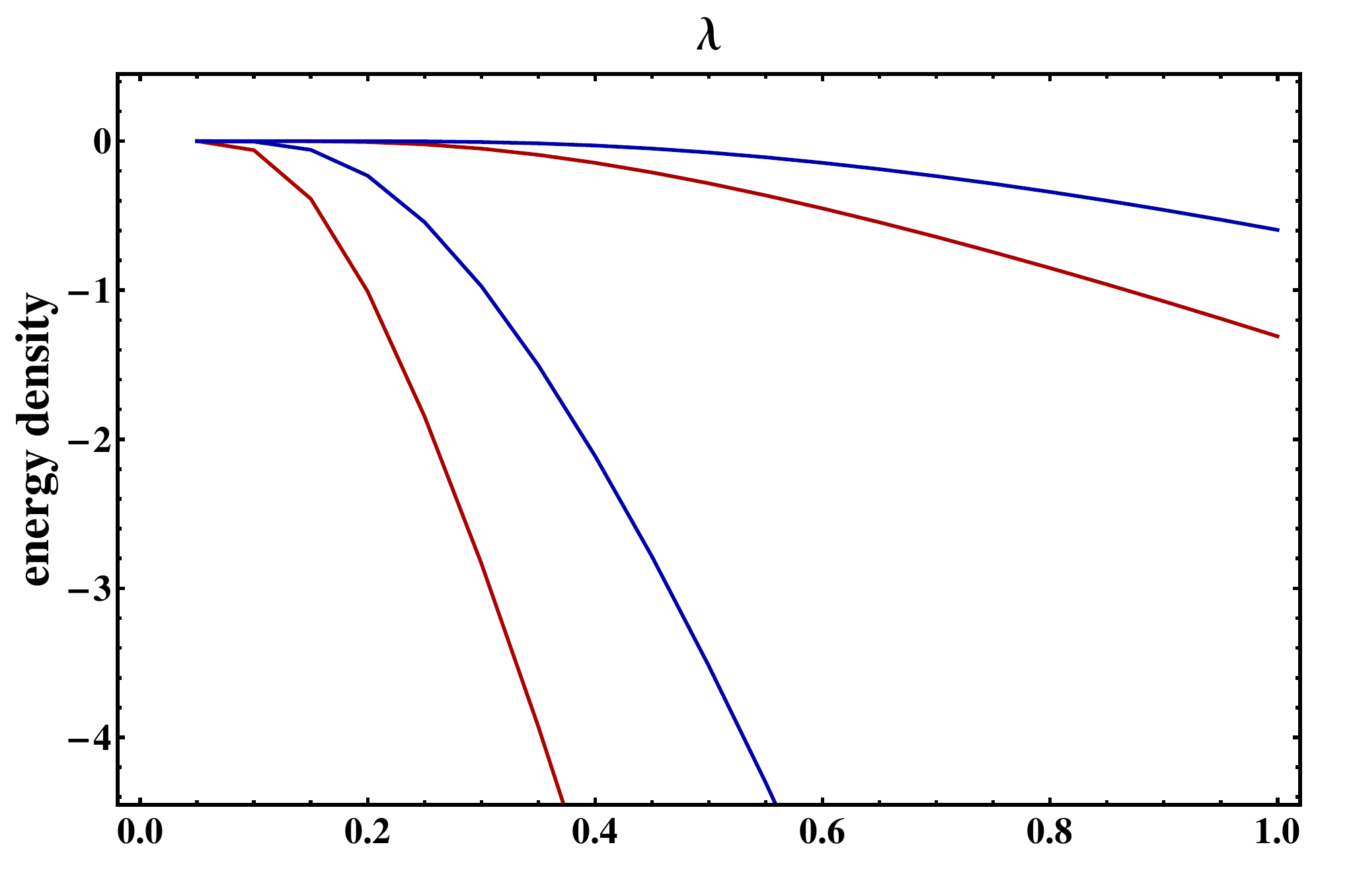}
\end{center}
\par
\vspace{-0.5cm}
\caption{Condensation energy in clean singlet (red) and triplet (blue)
superconductor as function of the phonon mediated effective
electron-electron coupling $\protect\lambda $ at two values of chemical
potential: $\protect\mu =8T_{D}$ (left),$\ 12T_{D}$ (right).}
\end{figure}
Magnetic impurities can strongly affect the relative energy of triplet and
especially singlet condensates for which it is pair breaking, see Fig.1.

\textit{Magnetic impurities.} Hamiltonian for impurity spins $\mathbf{s}_{a}$
located at $\mathbf{r}_{a}$ is

\begin{equation}
H_{imp}=-J\sum\nolimits_{a}\int_{\mathbf{r}}\delta \left( \mathbf{r}-\mathbf{%
r}_{a}\right) \mathbf{s}_{a}\cdot \mathbf{S}\left( \mathbf{r}\right) \text{,}
\label{Himp}
\end{equation}%
where itinerary spin density $\mathbf{S}$ was defined in Eq.(\ref{spin}).
Spins are randomly distributed:

\begin{equation}
\left\langle s_{a}^{i}s_{b}^{j}\delta \left( \mathbf{r}-\mathbf{r}%
_{a}\right) \delta \left( \mathbf{r}^{\prime }-\mathbf{r}_{b}\right)
\right\rangle _{dis}=\frac{s\left( s+1\right) }{3}n\delta ^{ij}\delta \left(
\mathbf{r}-\mathbf{r}^{\prime }\right) \text{,}  \label{distribution}
\end{equation}%
where $n$ is density of impurities and $s\hbar $ - their spin value.

Due to disorder the singlet, predictably gains energy over triplet and at
certain disorder strength a phase transition from the singlet to triplet
takes place. At yet large disorder strength the singlet channel becomes
unstable and the triplet becomes the only stable channel. As will be shown
below the triplet channel is generally not destabilized by this type of
impurities, so it is not a pair breaking.\bigskip\ After averaging over
impurities, see SM, the Gor'kov equations, Eq.(\ref{Gorkov}) acquires an
additional term dependent on frequency $\omega $ for the singlet:
\begin{equation}
L_{imp}^{+}\left( \omega \right) =-C\sum\nolimits_{\mathbf{q}}\Sigma
^{it}F^{+}\left( \mathbf{q,}\omega \right) \Sigma ^{i}\equiv -\frac{3iC}{g}%
\Delta _{S}\left( \omega \right) \alpha _{y}\text{.}  \label{Limp}
\end{equation}%
The dimensionless disorder strength is
\begin{equation}
C=s\left( s+1\right) nJ^{2}T_{D}/3\hbar ^{3}v_{F}^{3}  \label{C}
\end{equation}
and matrices $\Sigma $ were defined in Eq.(\ref{spin}). At a critical
disorder strength where the singlet channel is suppressed, $\Delta _{S}=0$
(so that $G\approx D$), the Gorkov equation integrated over momenta takes a
form%
\begin{equation}
\Delta _{S}\left( \omega \right) =f\left( \omega \right) \left( -3C\Delta
_{S}\left( \omega \right) +g\Delta _{S}\right) \text{,}  \label{Gorkocimp}
\end{equation}%
where $\Delta _{S}=\sum\nolimits_{\mathbf{q}}\Delta \left( \omega \right) $
and;

\begin{eqnarray}
f\left( \omega \right)  &=&\frac{1}{4}\text{tr}\sum\limits_{q}D^{t}\alpha
_{y}D\alpha _{y}  \label{f} \\
&=&\sum\limits_{q}\frac{\omega ^{2}+\mu ^{2}+v^{2}q^{2}}{\left(
v^{2}q^{2}+\omega ^{2}-\mu ^{2}\right) ^{2}+4\omega ^{2}\mu ^{2}}\text{.}
\notag
\end{eqnarray}%
To solve the equations for the critical disorder strength $C^{c}$, one
integrates over $\omega $:

\begin{equation}
\frac{1}{g}=\sum\nolimits_{\omega }\frac{f\left( \omega \right) }{%
1+3C^{c}f\left( \omega \right) }\text{.}  \label{crit}
\end{equation}%
The phase diagram for chemical potential $\mu =4T_{D}$, $8T_{D}$, $12T_{D}$
in wide range of $\lambda $ and $C$ is presented in Fig.4. Above the line
there is no singlet condensate, while triplet is the ground state. Below the
line the singlet pairing exists and possible dominates over the triplet. For
the triplet pairing calculation one obtains an equation similar to Eq.(\ref%
{crit}) with reverse sign in denominator and no solution. This means that
magnetic impurities help the pairing rather than destroying it.
\begin{figure}[tbp]
\begin{center}
\includegraphics[width=8cm]{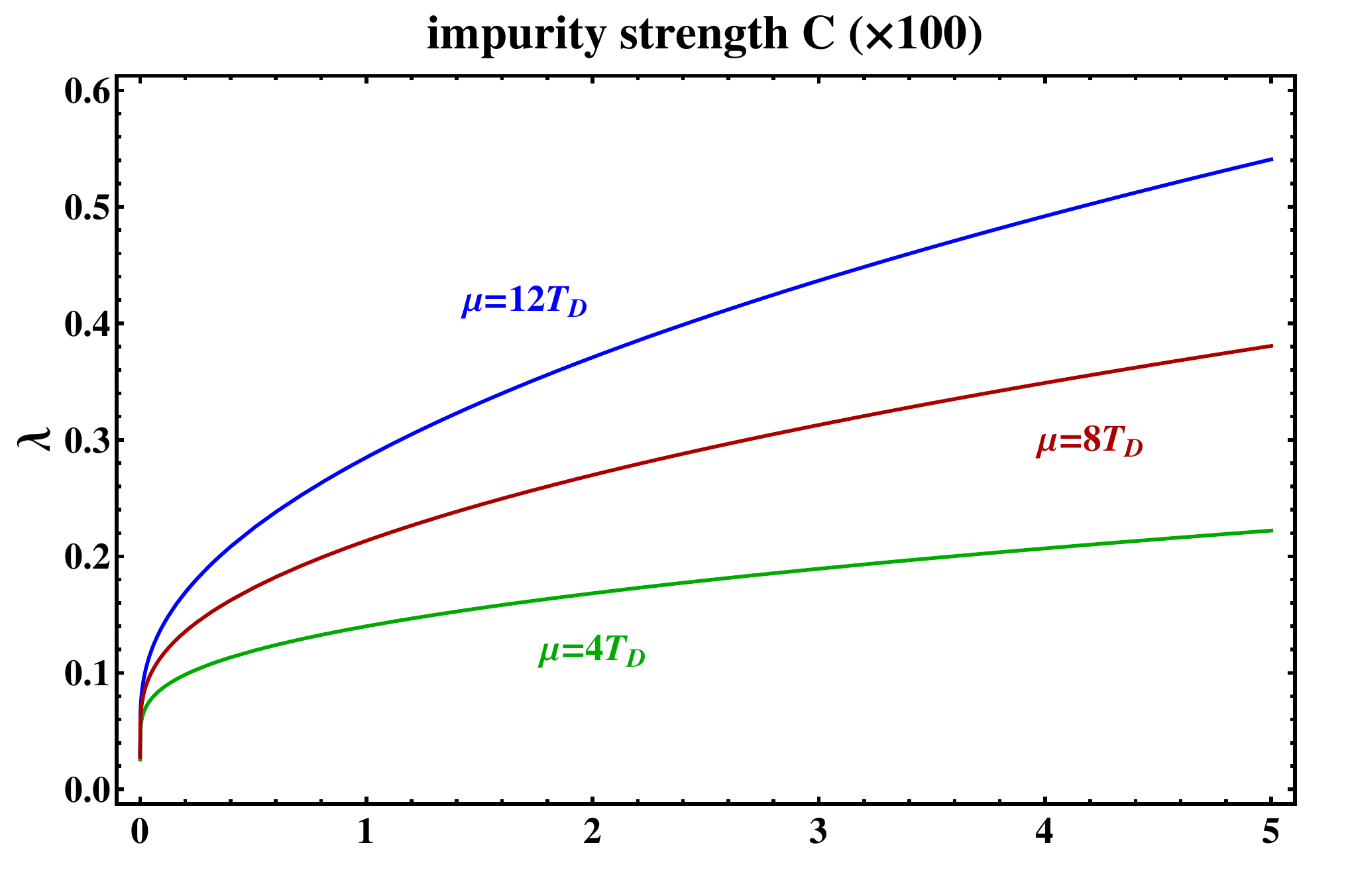}
\end{center}
\par
\vspace{-0.5cm}
\caption{Phase diagram of the magnetically doped 3D Dirac
semimetal in the $\protect\lambda -C$ plane for different chemical
potentials.\textit{\ } $\protect\lambda $ is the phonon mediated
electron-electron coupling, while $C$ is the magnetic impurities strength
defined via concentration and the in Eq.(\protect\ref{C}).}
\end{figure}

\textit{Summary and discussion. }To summarize, we presented a microscopic
theory of superconductivity (at zero temperature) in massless Dirac semi -
metals. In the framework of the "conventional" phonon mediated local
attraction model we classified (under simplifying assumptions of the 3D
rotation invariance, inversion and the time reversal) possible pairing
channels. There are three even parity (singlet) channels and one odd parity
(triplet) channels. In the clean limit the singlet pairing prevails for the
arbitrary chemical potential and the electron-electron interaction strength
despite the fact that triplet condensate is sometimes higher. This is found
by the direct comparison of condensation energies. However a modest
concentration of magnetic impurities makes the triplet ground state. Larger
impurity concentration suppresses the even parity state all together, while
the impurities are not pair breaking for the odd parity weakly ferromagnetic
state, Fig.1.

Here we compare our results with the early work ref.\cite{FuBerg} designed
to model the symmetries and parameters of Cu doped $Bi_{2}Se_{3}$. The case
that can be directly compared is when the relativistic mass term (denoted by
$m$ in ref.\cite{FuBerg}) is small compared to chemical potential. In this
work more general effective electron - electron interaction was considered
with two couplings $V$ and $U$ for local intraband and interband attractions
respectively. They are related to our $g$ by $g=2U=2V$. Qualitatively indeed
for $U/V=1$ one gets nearly degenerate energies (critical temperatures were
compared in ref.\cite{FuBerg}\ instead). This is similar but not identical
to our result without impurities. We indeed obtain the near degeneracy of
the two gaps, the singlet and the triplet (their $\Delta _{1}$ and $\Delta
_{2}$ respectively), but only in the limit of large $g$. The gaps are
definitely not degenerate when the coupling $g$ is below $20\pi
^{2}v_{F}^{3}\hbar ^{2}/T_{D}^{2}$. Even within the BCS regime (SM), $\Delta
_{T}/\Delta _{S}=\sinh \left( 0.35/\lambda \right) /\sinh \left( 0.5/\lambda
\right) $. This is consistent with $1$ only for quite large coupling and was
studied in detail in \cite{arxiv}.

To estimate the range of parameters for currently available materials where
the odd parity conventional (phonon induced) superconductivity is expected,
one should rely on measurements of the electron - phonon coupling. The
effective dimensionless electron - electron coupling constant due to phonons
$\lambda $ for materials like $Bi_{2}Se_{3}$, $Bi_{2}Te_{3}$ reported \cite%
{Howard} vary widely $0.1-3$. Taking\cite{phononexp} for $Bi_{2}Se_{3}$ the
Debye cutoff energy $T_{D}=150K$ and $\lambda =0.2$ measured at $\mu =120$ $%
meV$, Fermi velocity $7\cdot 10^{7}cm/s$ one obtains $6K$ triplet
superconductivity (see Fig.2 and a stronger singlet). To destroy the singlet
one, that is to reach the impurity strength $C=0.005$ that for the impurity
spin $\,s=1$ and exchange integral of $J=0.15$ $eV\cdot nm^{3}$ requires the
impurity concentration $n=2\cdot 10^{21}cm^{-3}$.

The physics of the triplet superconductors of this type is very rich and has
already been investigated in connection with heavy fermion superconductors.
In particular their magnetic vortices appear as either vector vortices or so
called skyrmions\cite{Knigavko} - coreless topologically nontrivial
textures. In particular their magnetic properties like the magnetization are
very peculiar and even without magnetic field the system forms a
"spontaneous flux state". The material therefore can be called a
"ferromagnetic superconductor". The superconducting state develops weak
ferromagnetism and system of alternating magnetic domains\cite{Bel}.

\acknowledgments
 We are indebted to C. W. Luo, T. Maniv and M.
Lewkowicz for valuable discussions. Work of B.R. and D.L. was supported by
NSC of R.O.C. Grants No. 98-2112-M-009-014-MY3 and MOE ATU program. The work
of D.L. also is supported by National Natural Science Foundation of China
(No. 11274018).

\onecolumn 
\setcounter{equation}{0}
\renewcommand\theequation{S. \arabic{equation}} 
\setcounter{figure}{0}
\renewcommand\thefigure{S. \arabic{figure}} 

\begin{center}
\begin{large}
\textbf{Supplemental material}
\end{large}
\end{center}

\section{1. Symmetry classification of pairing channels}

Electrons in 3D Dirac semi-metal are described by a four component bi-spinor
creation operator, $\psi _{\alpha }^{\dagger }= \psi _{L\uparrow }^{\dagger
},\psi _{L\downarrow }^{\dagger },\psi _{R\uparrow }^{\dagger }$,\\
$\psi_{R\downarrow }^{\dagger }$, whose index $\gamma $ takes four values. Here
we classify the possible local superconducting order parameters, written
generally as
\begin{equation}
\widehat{M}=\int_{\mathbf{r}}\psi _{\alpha }^{+}\left( \mathbf{r}\right)
M_{\alpha \beta }\psi _{\beta }^{+}\left( \mathbf{r}\right) \text{,}
\label{SA1a}
\end{equation}%
with constant antisymmetric matrix $M$ according to representations of the
3D rotation group. The representations of the rotation group therefore
characterize various possible superconducting phases. Generator of rotations
consists of the orbital momentum operator $\mathbf{L}$ and the spin operator%
\begin{equation}
S^{i}=\int_{\mathbf{r}}\psi _{\gamma }^{+}\left( r\right) \mathbf{\Sigma }%
_{\gamma \delta }^{i}\psi _{\delta }\left( r\right) ,  \label{SA1b}
\end{equation}%
Due to the rotation symmetry they transform covariantly under the action of $%
\mathbf{J}=\mathbf{L}+\mathbf{S}$. The global quantity in Eq.(\ref{SA1a})
transforms as

\begin{eqnarray}
\left[ \widehat{M},J^{i}\right] &=&\int_{\mathbf{r,r}^{\prime }}\left[ \psi
_{\alpha }^{+}\left( r\right) M_{\alpha \beta }\psi _{\beta }^{+}\left(
r\right) ,\psi _{\gamma }^{+}\left( r^{\prime }\right) \mathbf{\Sigma }%
_{\gamma \delta }^{i}\psi _{\delta }\left( r^{\prime }\right) \right]
\label{transformation} \\
&=&-2\int_{r}\psi _{\gamma }^{+}\left( r\right) \mathbf{\Sigma }_{\gamma
\delta }^{i}M_{\delta \kappa }\psi _{\kappa }^{+}\left( r\right) \text{.}
\notag
\end{eqnarray}%
Out of 16 possible matrices $M$ six are antisymmetric. They transform into
each other forming the following irreducible representations.

(i) a triplet of matrices $\left\{ T_{x},T_{y},T_{z}\right\} =\left\{ \beta
\alpha _{z},-\gamma _{x}\gamma _{y}\gamma _{z},\beta \alpha _{x}\right\} $
transforms as a vector

\begin{equation}
\left[ \widehat{M_{k}^{T}},J^{l}\right] =i\varepsilon _{klm}\widehat{%
M_{m}^{T}}  \label{vector}
\end{equation}
\ \ \ \ \ \ \ \ (ii) three scalar multiplets: $S_{1}=i\alpha _{y};$ \ \ $%
S_{2}=i\Sigma _{y};$ \ \ $S_{3}=-i\beta \alpha _{y}\gamma _{5}$. \

Which one of the condensates is realized at zero temperature is determined
by the Hamiltonian.

\section{2. Microscopic equations for local pairing}

\subsection{Gor'kov equations}

To treat the pairing the general gaussian approximation can be employed.
Using the standard formalism, the Matsubara Green's functions ($\tau $ is
the Matsubara time),
\begin{eqnarray}
G_{\alpha \beta }\left( \mathbf{r},\tau ;\mathbf{r}^{\prime },\tau ^{\prime
}\right) &=&-\left\langle T_{\tau }\psi _{\alpha }\left( \mathbf{r},\tau
\right) \psi _{\beta }^{\dagger }\left( \mathbf{r}^{\prime },\tau ^{\prime
}\right) \right\rangle \text{;}  \label{GFdef} \\
F_{\alpha \beta }^{\dagger }\left( \mathbf{r},\tau ;\mathbf{r}^{\prime
},\tau ^{\prime }\right) &=&\left\langle T_{\tau }\psi _{\alpha }^{\dagger
}\left( \mathbf{r},\tau \right) \psi _{\beta }^{\dagger }\left( \mathbf{r}%
^{\prime },\tau ^{\prime }\right) \right\rangle \text{,}  \notag
\end{eqnarray}%
obey the Gor'kov equations:%
\begin{gather}
-\frac{\partial G_{\gamma \kappa }\left( \mathbf{r},\tau ;\mathbf{r}^{\prime
},\tau ^{\prime }\right) }{\partial \tau }-\int_{\mathbf{r}^{\prime \prime
}}\left\langle \mathbf{r}\left\vert \widehat{K}_{\gamma \beta }\right\vert
\mathbf{r}^{\prime \prime }\right\rangle G_{\beta \kappa }\left( \mathbf{r}%
^{\prime \prime },\tau ;\mathbf{r}^{\prime },\tau ^{\prime }\right)
\label{Gorkov} \\
-gF_{\beta \gamma }\left( \mathbf{r},\tau ;\mathbf{r},\tau \right) F_{\beta
\kappa }^{\dagger }\left( \mathbf{r},\tau ,\mathbf{r}^{\prime },\tau
^{\prime }\right) =\delta ^{\gamma \kappa }\delta \left( \mathbf{r-r}%
^{\prime }\right) \delta \left( \tau -\tau ^{\prime }\right) ;  \notag \\
\frac{\partial F_{\gamma \kappa }^{\dagger }\left( \mathbf{r},\tau ;\mathbf{r%
}^{\prime },\tau ^{\prime }\right) }{\partial \tau }-\int_{\mathbf{r}%
^{\prime \prime }}\left\langle \mathbf{r}\left\vert \widehat{K}_{\gamma
\beta }^{t}\right\vert \mathbf{r}^{\prime \prime }\right\rangle F_{\beta
\kappa }^{\dagger }\left( \mathbf{r}^{\prime \prime },\tau ;\mathbf{r}%
^{\prime },\tau ^{\prime }\right)  \notag \\
-gF_{\gamma \beta }^{\dagger }\left( \mathbf{r},\tau ;\mathbf{r},\tau
\right) G_{\beta \kappa }\left( \mathbf{r},\tau ,\mathbf{r}^{\prime },\tau
^{\prime }\right) =0\text{.}  \notag
\end{gather}%
In the homogeneous case the Gor'kov equations for Fourier components of the
Greens functions simplify considerably,
\begin{eqnarray}
D_{\gamma \beta }^{-1}G_{\beta \kappa }\left( \omega ,p\right) -\Delta
_{\gamma \beta }F_{\beta \kappa }^{\dagger }\left( \omega ,p\right)
&=&\delta ^{\gamma \kappa }\text{;}  \label{Gorkov_uniform} \\
D_{\beta \gamma }^{-1}F_{\beta \kappa }^{\dagger }\left( \omega ,p\right)
+\Delta _{\gamma \beta }^{\ast }G_{\beta \kappa }\left( \omega ,p\right) &=&0%
\text{,}  \notag
\end{eqnarray}%
where $\omega =\pi T\left( 2n+1\right) $ is the Matsubara frequency and$\
D_{\gamma \beta }^{-1}=\left( i\omega -\mu \right) \delta _{\gamma \beta
}+v_{F}p^{j}\alpha _{\alpha \beta }^{j}$.

\bigskip The matrix gap function can be chosen as ($\Delta $ real)
\begin{equation}
\widehat{\Delta }_{\beta \gamma }=gF_{\gamma \beta }\left( 0\right) =\Delta
M_{\gamma \beta }\text{.}  \label{delta}
\end{equation}%
These equations are conveniently presented in matrix form (superscript $t$
denotes transposed and $I$ - the identity matrix):
\begin{eqnarray}
D^{-1}G-\Delta F^{\dagger } &=&I\text{;}  \label{matrixeq} \\
D^{t-1}F^{\dagger }+\Delta ^{\ast }G &=&0\text{.}  \notag
\end{eqnarray}%
Solving these equations, one obtains
\begin{eqnarray}
G^{-1} &=&D^{-1}+\Delta D^{t}\Delta ^{\ast }\text{;}  \label{solution} \\
F^{\dagger } &=&-D^{t}\Delta ^{\ast }G\text{,}  \notag
\end{eqnarray}%
with the gap function, Eq.(\ref{gap}), found from the consistency condition.
Now we find solutions of this equation for each of the possible
superconducting phases.

\subsection{Triplet solution of the gap equation}

In this phase rotational symmetry is spontaneously broken simultaneously
with the electric charge $U\left( 1\right) $ (global gauge invariance)
symmetry. Assuming $z$ direction of the $p$ - wave condensate the order
parameter matrix takes a form: $\Delta =\Delta _{T}M_{z}^{T}=\Delta
_{T}\beta \alpha _{x}$. In this Section we use the units of $v_{F}=1,\hbar
=1 $ and the energy scale will be set by the Debye cutoff, $T_{D}=1$, of the
electron - phonon interactions, see below. The off-diagonal matrix element
of the matrix gap equation, for real $\Delta _{T}>0$ is:
\begin{equation}
\frac{1}{g}=\sum\limits_{\omega q}\frac{\Delta _{T}^{2}+p_{\perp
}^{2}-p_{z}^{2}+\mu ^{2}+\omega ^{2}}{\left( \Delta _{T}^{2}+\omega
^{2}\right) ^{2}+\left( p^{2}-\mu ^{2}\right) ^{2}+2\left( p^{2}+\mu
^{2}\right) \omega ^{2}+2\Delta _{T}^{2}\left( p_{\perp }^{2}-p_{z}^{2}+\mu
^{2}\right) }\text{,}  \label{gap}
\end{equation}%
where $p_{\perp }^{2}=p_{x}^{2}+p_{y}^{2}$. The spectrum of elementary
excitations obtained from the four poles of the Greens function, see
Fig.SM1, is\ (in physical units)
\begin{equation}
E_{\pm }^{2}=\Delta _{T}^{2}+v_{F}^{2}p^{2}+\mu ^{2}\pm 2v_{F}\sqrt{\Delta
_{T}^{2}p_{z}^{2}+p^{2}\mu ^{2}}\text{.}  \label{spectrum}
\end{equation}%
There are two nodes at $p_{x}=p_{y}=0,v_{F}p_{z}=\pm \sqrt{\Delta
_{T}^{2}+\mu ^{2}}$, when the branches $+\left\vert E_{-}\right\vert $ and $%
-\left\vert E_{-}\right\vert $ cross, see Fig.SM1a and a section $p_{\perp
}=0$ in Fig.SM1b. There is also a saddle points with energy gap, $2\Delta
_{T}$ on the circle $p_{x}^{2}+p_{y}^{2}=\mu ^{2},p_{z}=0$ see the section
in the $p_{z}=0$ direction in Fig. SM1c. The higher energy band $E_{+}$
touches the lower band at $p=0$, so that there is a Dirac point for
quasiparticles, see Fig. SM1d.

\begin{figure}[tbp]
\centering
\subfigure[]{\includegraphics[width=6cm]{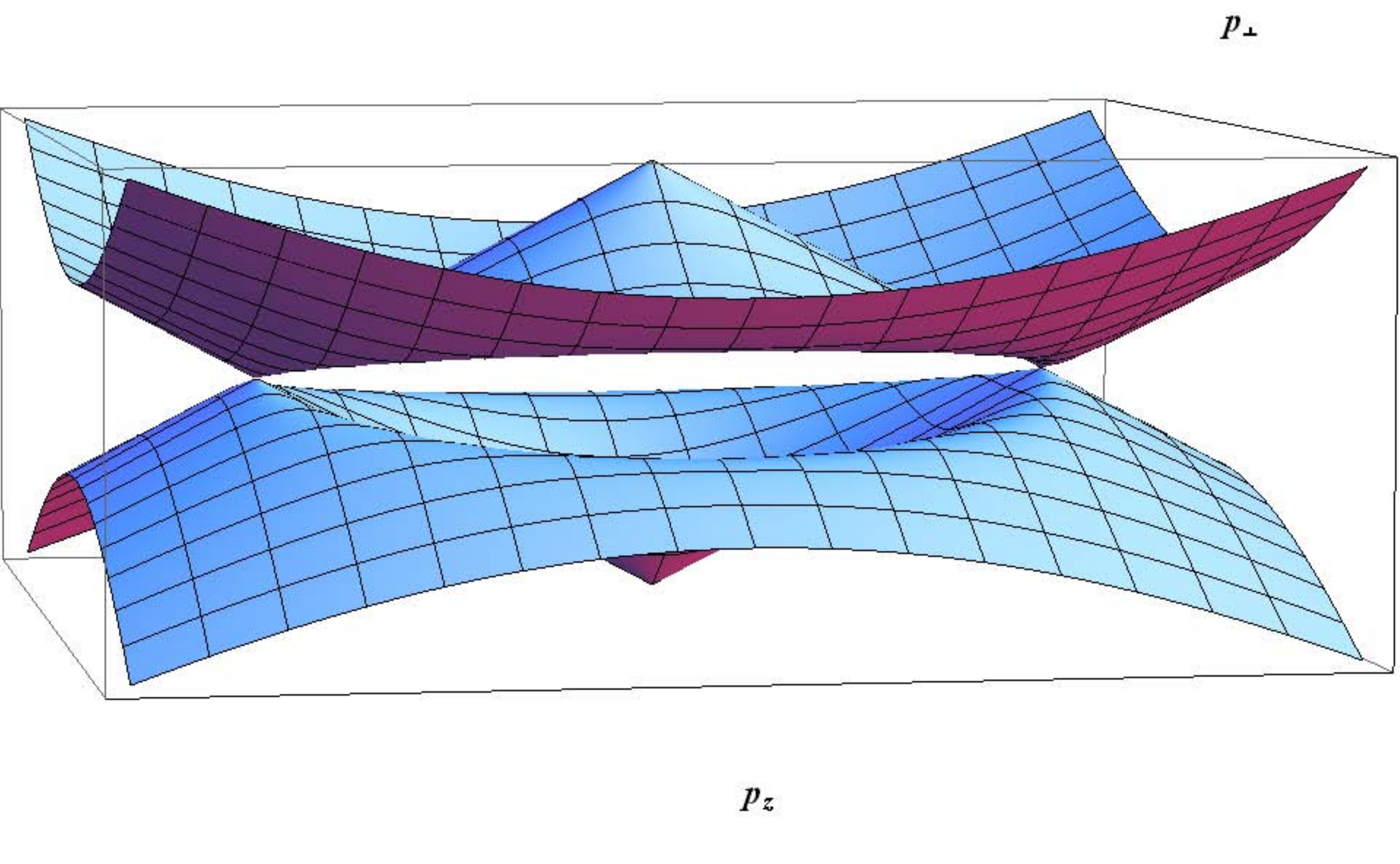}} 
\subfigure[]{\includegraphics[width=6cm]{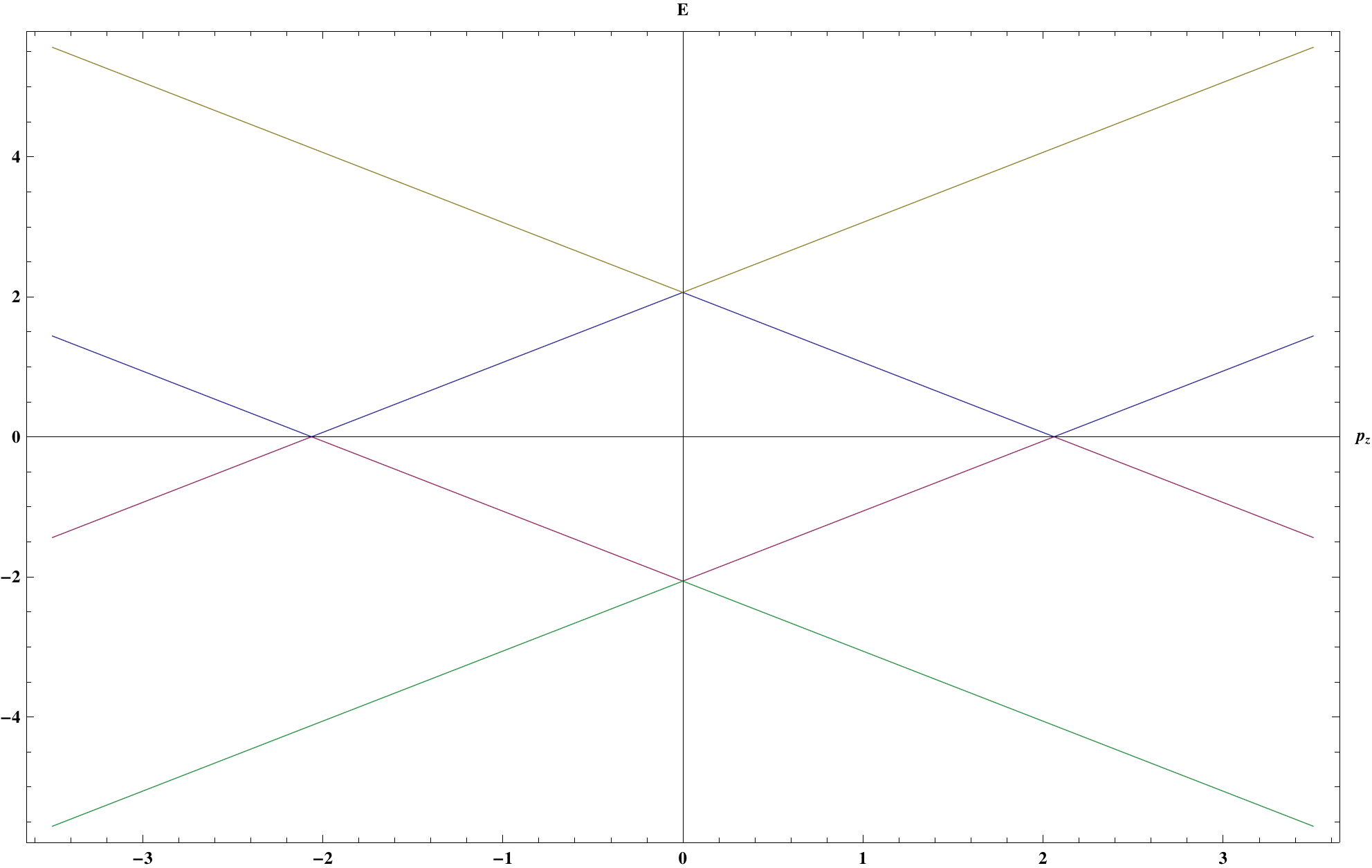}} 
\subfigure[]{\includegraphics[width=6cm]{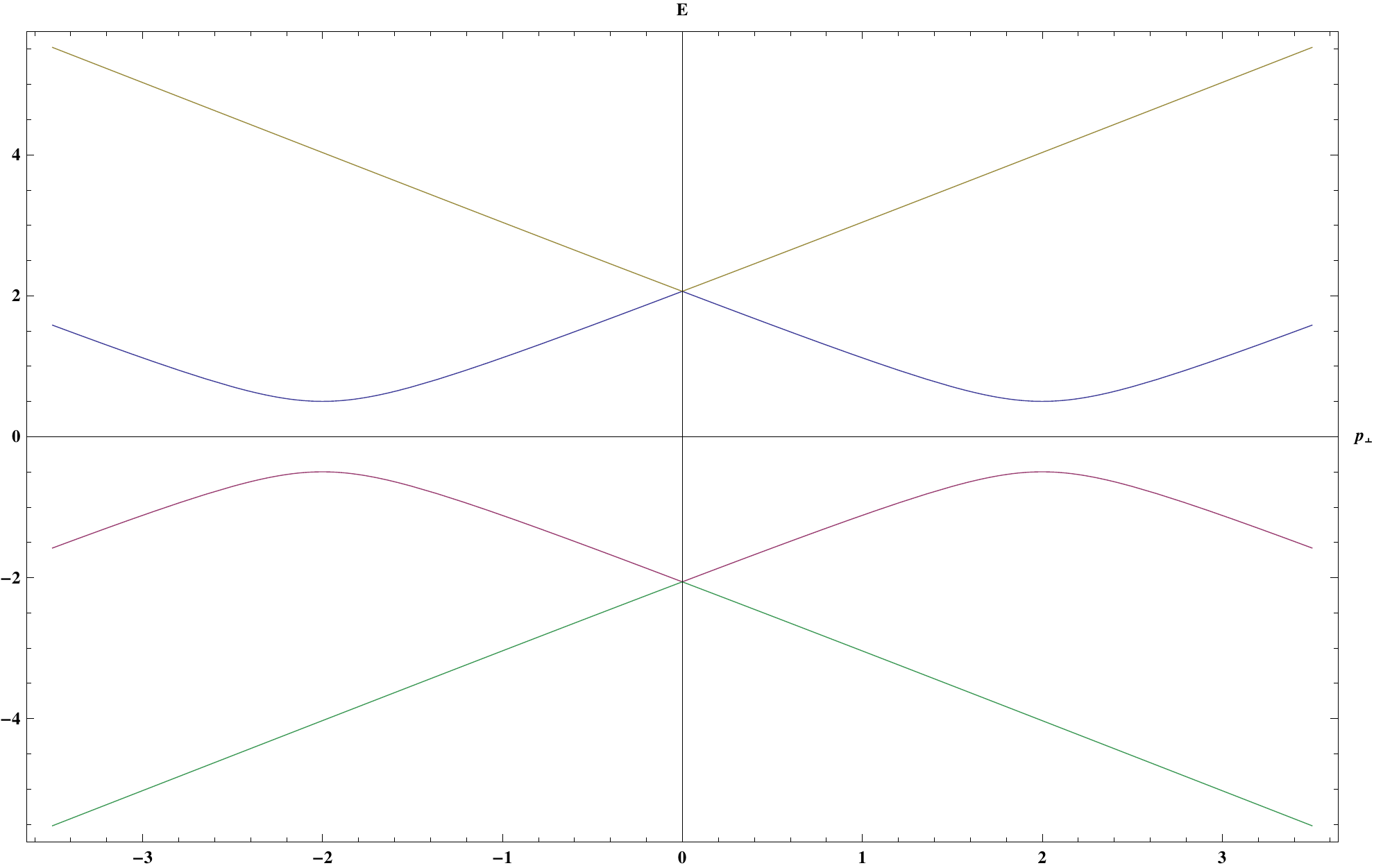}}
\subfigure[]{\includegraphics[width=6cm]{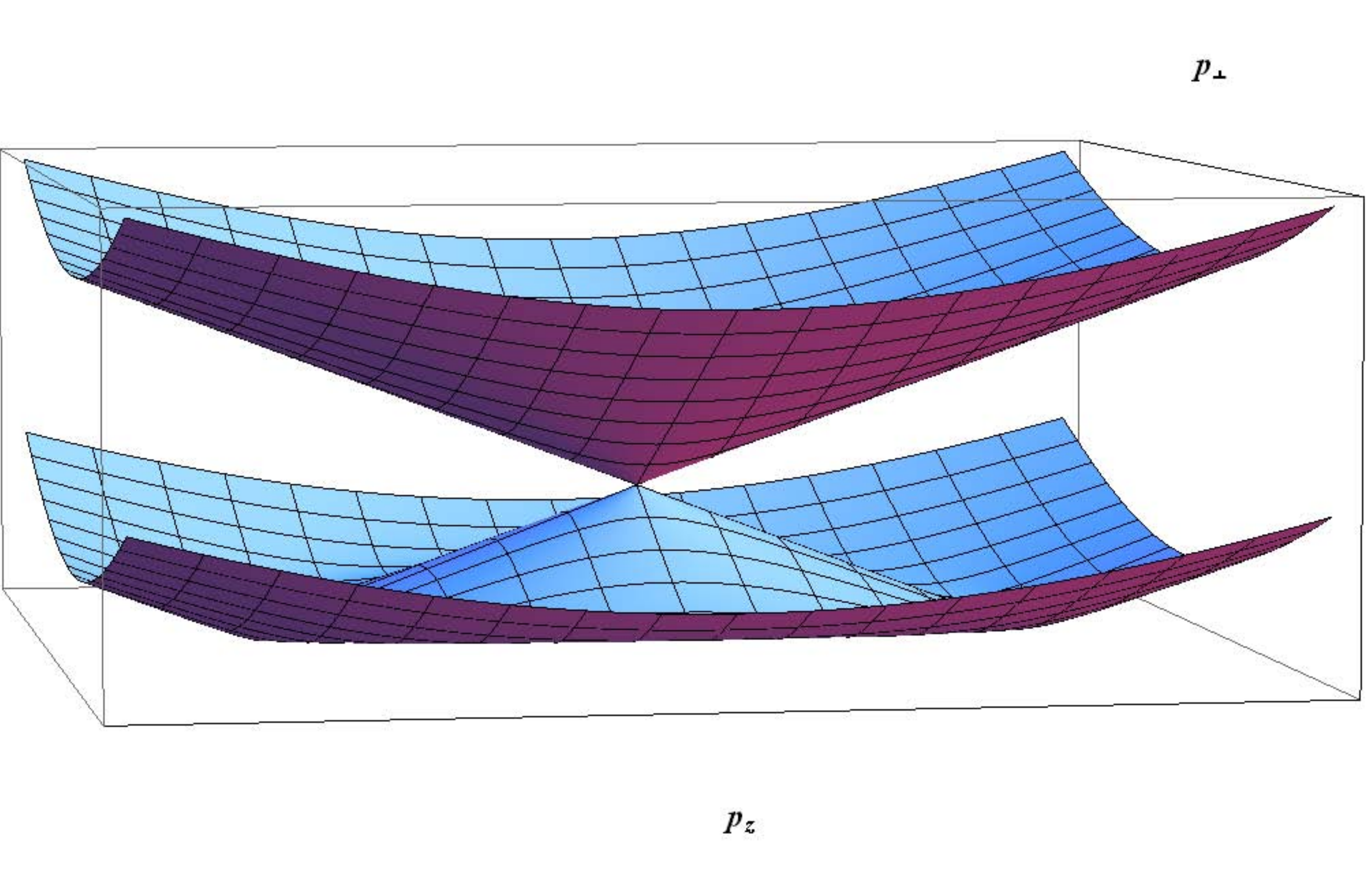}}
\caption{Spectrum of triplet excitations. a. section $p_{\perp }=0$ in b.
There is also a saddle points with energy gap, c. $2\Delta _{T}$ on the
circle $p_{x}^{2}+p_{y}^{2}=\protect\mu ^{2},p_{z}=0$ see the section in the
$p_{z}=0$ direction in . d. The higher energy band $E_{+}$ touches the lower
band at $p=0$, so that there is a Dirac point for quasiparticles.}
\end{figure}

Integration over $\omega $ gives using polar coordinates for $p$ and $x=\cos
\theta ,$ $\zeta =\sqrt{\Delta _{T}^{2}x^{2}+\mu ^{2}}$,%
\begin{equation}
\frac{1}{g}=\frac{1}{8\pi ^{2}}\int_{p=\max \left[ \mu -1,0\right] }^{\mu
+1}\int_{x=0}^{1}\frac{p^{2}}{\zeta }\left\{ \frac{\zeta +px^{2}}{\sqrt{%
\Delta _{T}^{2}+p^{2}+\mu ^{2}+2p\zeta }}+\frac{\zeta -px^{2}}{\sqrt{\Delta
_{T}^{2}+p^{2}+\mu ^{2}-2p\zeta }}\right\} \text{.}  \label{gapeqT}
\end{equation}%
The lower bound on the momentum integration is nonzero when chemical
potential $\mu $ exceeds $T_{D}$, see Fig. SM2. The integral over $x$ was
performed analytically, while the last integral was done numerically.

\begin{figure}[tbp]
\centering
\subfigure[]{\includegraphics[width=6cm]{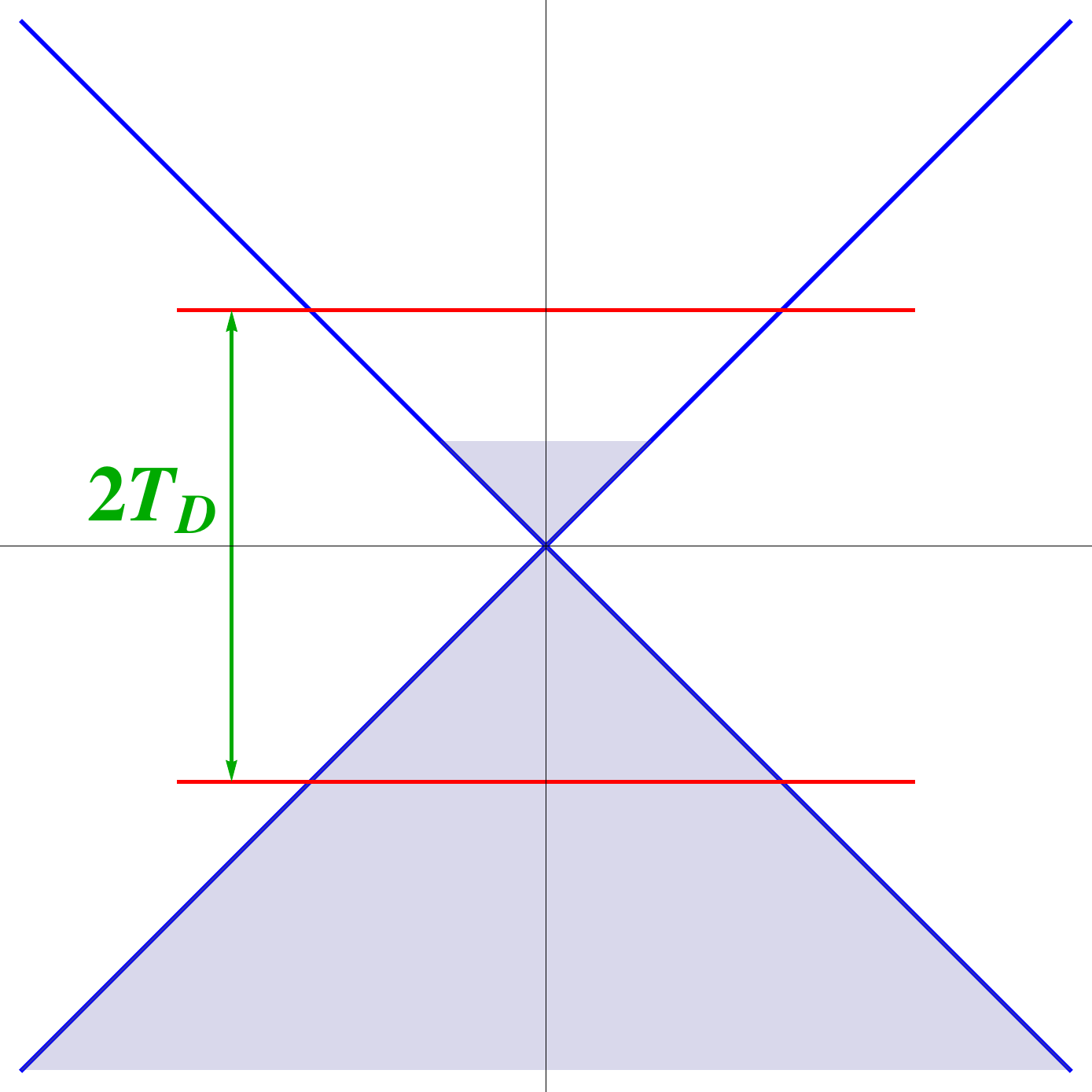}} \subfigure[]{%
\includegraphics[width=6cm]{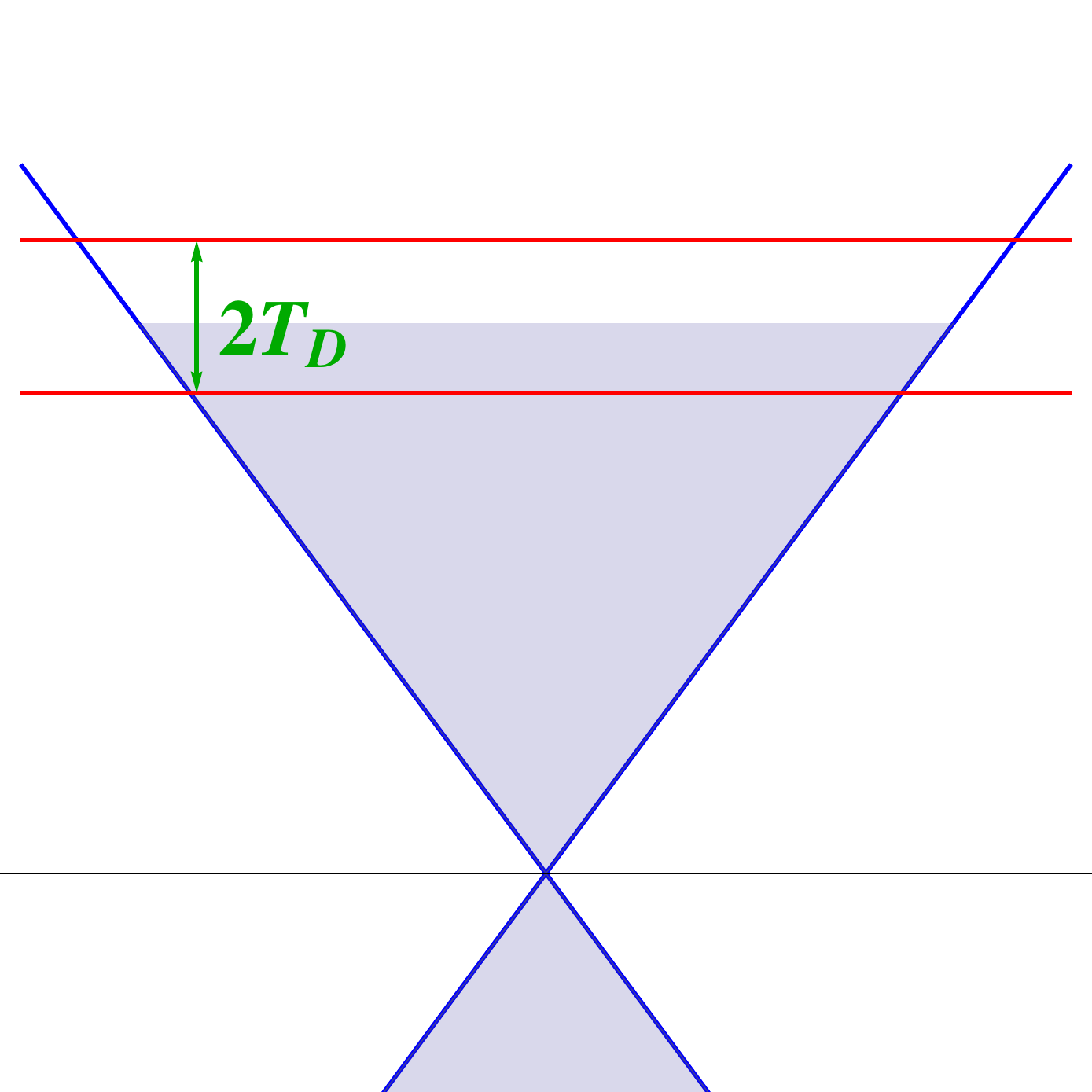}}
\caption{Chemical potential in Dirac semi - metals and the phonon mediated
pairing. (a) Chemical potential relative to Dirac point is smaller that
typical energy of phonons, the Debye energy $T_{D}$. (b) The BCS
approximation limit: the chemical potential is much larger than the Debye
energy $T_{D}$.}
\end{figure}

\subsection{Singlet representations}

It turns out that the second singlet in Eq.(\ref{gapsinglet}) gives results
identical to that of the first one, while the third singlet does not have a
solution in the physically interesting range of parameters. Therefore we
assume the order parameter in the matrix form $\ \Delta =\Delta
_{S}M_{1}^{S}=i\Delta _{S}\alpha ^{y}$. The relevant matrix element of the
matrix gap equation, is for real $\Delta _{S}$:

\begin{equation}
\frac{1}{g}=\sum\limits_{\omega p}\frac{\Delta _{S}^{2}+p^{2}+\mu
^{2}+\omega ^{2}}{\left( \Delta _{S}^{2}+p^{2}\right) ^{2}+\left( \mu
^{2}+\omega ^{2}+2\Delta _{S}^{2}\right) \left( \mu ^{2}+\omega ^{2}\right)
+2p^{2}\left( \omega ^{2}-\mu ^{2}\right) }\text{.}  \label{gapsinglet}
\end{equation}%
Spectrum (in physical units) now is isotropic,%
\begin{equation}
E_{\pm }^{2}=\Delta _{S}^{2}+\left( v_{F}\left\vert p\right\vert \pm \mu
\right) ^{2}\text{.}  \label{spectrum_S}
\end{equation}%
Integration over $\omega $ gives%
\begin{equation}
\frac{1}{g}=\mu \sum\limits_{\mu -T_{D}<\varepsilon _{p}<\mu +T_{D}}\frac{p}{%
r_{+}r_{-}\left( r_{+}-r_{-}\right) }\text{,}  \label{gap1_S}
\end{equation}%
where $r_{\pm }=\sqrt{\Delta _{S}^{2}+\left( \left\vert p\right\vert \pm \mu
\right) ^{2}}$, while the $p$ integration results in:%
\begin{equation}
\frac{16\pi ^{2}}{g}=\Phi \left( \mu +1,\mu \right) -\Phi \left( \max \left[
\mu -1,0\right] ,\mu \right)  \label{gapeq2_S}
\end{equation}%
with%
\begin{equation}
\Phi \left( p,\mu \right) =r_{-}\left( p+3\mu \right) +r_{+}\left( p-3\mu
\right) -\left( \Delta _{S}^{2}-2\mu ^{2}\right) \log \left[ \left(
p+r_{-}-\mu \right) \left( p+r_{+}+\mu \right) \right]  \label{Fidef}
\end{equation}

The solution is presented in Fig. 2 of the paper as lines of constant
chemical potential. Having found the order parameter, one has to determine
what symmetry breaking is realized by comparing energies of the solutions
as explained in the text.

\section{3. The BCS and the strong coupling limits}

\subsection{Triplet}

In several limiting cases the integrals can be performed analytically. At
zero chemical potential the results are presented in Section IV, while here
we list the BCS limit of $\mu >>T_{D}$ and the strong coupling case of $g\mu
^{2}>>1$, $\Delta _{T}\propto g$.

(i) In the BCS limit one has%
\begin{equation}
\frac{1}{g}=\frac{a_{T}\mu ^{2}}{4\pi ^{2}}\sinh ^{-1}\frac{T_{D}}{\Delta
_{T}}\text{,}  \label{gapBCS_T}
\end{equation}%
with $a_{T}=0.69$, leading to exponential gap dependence on $\lambda $ when
it is small:%
\begin{equation}
\Delta _{T}=T_{D}/\sinh \left( 1/2a_{T}\lambda \right) \simeq
2T_{D}e^{-1/2a_{T}\lambda }\text{.}  \label{dT_BCS}
\end{equation}

(ii) In the strong coupling one obtains with solution
\begin{equation}
\Delta _{T}=\frac{g}{12\pi ^{2}}\left\{
\begin{array}{c}
6\mu ^{2}+2\text{ \ for }\mu <1 \\
\left( \mu +1\right) ^{3}\text{\ for }\mu >1%
\end{array}%
\right. \text{.}  \label{dT_sc}
\end{equation}%
Usually the local coupling does not prefer the triplet pairing and the
singlet channels of coupling are realized. We therefore turn to them.

\subsection{\protect\bigskip Singlet}

For singlet one has

(i) BCS, $\mu >>T_{D}$

\begin{equation}
\Delta _{S}=T_{D}/\sinh \left( 1/2\lambda \right) \simeq
2T_{D}e^{-1/2\lambda }\text{.}  \label{BCS_S}
\end{equation}

(ii) Strong coupling

\begin{equation}
\Delta _{S}=\frac{2\lambda \left( T_{D}+\mu \right) ^{3}}{3\mu ^{2}}\text{.}
\label{S_sc}
\end{equation}

\subsection{Energies}

In limiting cases, one obtains expressions in closed form.

(i) BCS, $\mu >T_{D}$, using Eq.(\ref{gapBCS_T}) and Eq.(\ref{dT_BCS}) for
the triplet and Eq.(\ref{BCS_S}) for the singlet, one has the energy density:%
\begin{equation}
F_{T,S}=-\frac{a_{T,S}\mu ^{2}T_{D}}{2\pi ^{2}v_{F}^{3}\hbar ^{3}}\left(
\sqrt{\Delta _{T}^{2}+T_{D}^{2}}-T_{D}\right) \simeq -\frac{a_{T,S}}{\pi ^{2}%
}\frac{\mu ^{2}T_{D}^{2}}{v_{F}^{3}\hbar ^{3}}\exp \left( -\frac{1}{%
a_{T,S}\lambda }\right) ,  \label{F_BCS}
\end{equation}%
with $a_{T}=0.69$, while $a_{S}=1$ and assuming $\lambda <<1$. The ratio of
the two phases gives

\begin{equation}
\frac{F_{T}}{F_{S}}=0.69e^{-0.45/\lambda }\text{.}  \label{ratio}
\end{equation}
:

\bigskip (ii) Strong coupling limit, using Eq.(\ref{dT_sc}) for triplet and
Eq.(\ref{S_sc}) for the singlet,

\begin{equation}
F_{T}=F_{S}=-\frac{1}{72\pi ^{4}v_{F}^{3}\hbar ^{3}}\left\{
\begin{array}{c}
4\left( 3\mu ^{2}+T_{D}^{2}\right) ^{2}\text{ \ for }\mu <T_{D} \\
T_{D}^{-2}\left( \mu +T_{D}\right) ^{6}\text{\ \ for }\mu >T_{D}%
\end{array}%
\right. \text{.}  \label{energy_sc}
\end{equation}%
The difference appears at order $1/g$. To summarize, in most of the
parameter range shown triplet is a bit higher than that of the singlet, but
the two condensates are nearly degenerate.

\section{ 4. Magnetic impurities}

After averaging over impurities, the Gor'kov equations, Eq.(\ref%
{Gorkov}) acquires an additional term In components (no Nambu notations)%
\begin{eqnarray}
I &=&D^{-1}G-NG-LF^{+}  \label{S4a} \\
0 &=&\left( D^{t-1}-N^{+}\right) F^{+}-\left( L^{+}-\Delta ^{\ast }\right) G
\notag
\end{eqnarray}%
where the normal disorder average

\begin{eqnarray}
N^{\alpha \beta ^{\prime }}\left( r-r^{\prime },\tau -\tau ^{\prime }\right)
&=&J^{2}\left\langle \sum\nolimits_{a,b}S_{a}^{i}S_{b}^{i}\delta \left(
r-r_{a}\right) \delta \left( r^{\prime }-r_{b}\right) \right\rangle
_{dis}\times   \label{N} \\
\Sigma _{\alpha \beta }^{i}\Sigma _{\alpha ^{\prime }\beta ^{\prime
}}^{i^{\prime }}\left\langle T\psi _{\beta }\left( r,\tau \right) \psi
_{\alpha ^{\prime }}^{+}\left( r^{\prime },\tau ^{\prime }\right)
\right\rangle  &=&-C\delta \left( r-r^{\prime }\right) \Sigma _{\alpha \beta
}^{i}G_{\beta \alpha ^{\prime }}\left( 0,\tau -\tau ^{\prime }\right) \Sigma
_{\alpha ^{\prime }\beta ^{\prime }}^{i}  \notag
\end{eqnarray}%
lead to the renormalization of the chemical potential and relaxation time
that can be safely neglected for our purposes. The second, anomalous
disorder average%
\begin{eqnarray}
L_{imp}^{+\beta \beta ^{\prime }}\left( r-r^{\prime },\tau -\tau ^{\prime
}\right)  &=&-J^{2}\left\langle \sum\nolimits_{a,b}S_{a}^{i}S_{b}^{i}\delta
\left( r-r_{a}\right) \delta \left( r^{\prime }-r_{b}\right) \right\rangle
_{dis}\times   \label{L} \\
\Sigma _{\alpha \beta }^{i}\Sigma _{\alpha ^{\prime }\beta ^{\prime
}}^{i}\left\langle T\psi _{\alpha }^{+}\left( r,\tau \right) \psi _{\alpha
^{\prime }}^{+}\left( r^{\prime },\tau ^{\prime }\right) \right\rangle
&=&-C\delta \left( r-r^{\prime }\right) \Sigma ^{it}F^{+}\left( 0,\tau -\tau
^{\prime }\right) \Sigma ^{i}\text{,}  \notag
\end{eqnarray}%
determines the influence of the disorder on the condensate. For singlet one
has in Fourier space%
\begin{equation}
\sum\nolimits_{q}F_{\beta \gamma }^{+}\left( q,\omega \right) =\frac{i}{g}%
\Delta ^{S}\left( \omega \right) \alpha _{\beta \gamma }^{y};  \label{Fomega}
\end{equation}%
leading via
\begin{equation}
i\Sigma ^{it}\alpha ^{y}\Sigma ^{i}=i\left(
\begin{array}{cc}
\sigma _{i}^{t} &  \\
& \sigma _{i}^{t}%
\end{array}%
\right) \left(
\begin{array}{cc}
\sigma _{y} &  \\
& -\sigma _{y}%
\end{array}%
\right) \left(
\begin{array}{cc}
\sigma _{i} &  \\
& \sigma _{i}%
\end{array}%
\right) =-3i\alpha _{y}  \label{algebra}
\end{equation}%
to Eq.(L) in the main text from which the bifurcation point is found.

Similarly for triplet
\begin{equation}
L_{imp}^{+}\left( p,\omega \right) =-C\Sigma
^{it}\sum\nolimits_{q}F^{+}\left( q,\omega \right) \Sigma ^{i}=-\frac{C}{g}%
\Delta ^{T}\left( \omega \right) \gamma ^{x}\text{.}  \label{tripletL}
\end{equation}%
since now%
\begin{equation}
\sum\nolimits_{q}F_{\beta \gamma }^{+}\left( q,\omega \right) =\frac{1}{g}%
\Delta ^{T}\left( \omega \right) \gamma _{\beta \gamma }^{x}
\label{Ftriplet}
\end{equation}%
and%
\begin{equation}
\Sigma ^{it}\gamma ^{x}\Sigma ^{i}=\left(
\begin{array}{cc}
\sigma _{i}^{t} & 0 \\
0 & \sigma _{i}^{t}%
\end{array}%
\right) \left(
\begin{array}{cc}
0 & -\sigma _{x} \\
\sigma _{x} & 0%
\end{array}%
\right) \left(
\begin{array}{cc}
\sigma _{i} & 0 \\
0 & \sigma _{i}%
\end{array}%
\right) =\gamma ^{x}  \label{algtriplet}
\end{equation}%
Note opposite signs of the singlet and triplet. At bifurcation point
(destruction of the condensate) the function $f$ will be now%
\begin{equation}
f_{T}\left( \omega \right) =\frac{1}{4}\text{tr}\sum\nolimits_{\mathbf{q}%
}D^{t}\gamma ^{x}D\gamma ^{x}>0\text{.}  \label{fT}
\end{equation}%
To solve the equations for the critical disorder strength $C^{c}$, one
integrates over $\omega $,

\begin{equation}
\frac{1}{g}=\sum\nolimits_{\omega }\frac{f_{T}\left( \omega \right) }{%
1-C^{c}f_{T}\left( \omega \right) }  \label{ft}
\end{equation}%
that has no solution.

\end{document}